\documentclass[twocolumn,showpacs,preprintnumbers,amsmath,amssymb]{revtex4}
\usepackage{graphicx}
\usepackage{bm}
\usepackage{float}
\usepackage[squaren]{SIunits}

\begin{document}
\title{Multiferroicity in the frustrated spinel cuprate GeCu$_{2}$O$_{4}$}

\author{L. Zhao, L. Muzica, U. Schwarz, and A. C. Komarek} \email{Alexander.Komarek@cpfs.mpg.de}
\affiliation{
Max-Planck-Institute for Chemical Physics of Solids, 01187 Dresden, Germany
}

\date{\today}

\begin{abstract}

Different from other magnetically frustrated spinel systems, GeCu$_{2}$O$_{4}$ is a strongly tetragonal distorted spinel cuprate in which edge-sharing CuO$_{2}$ ribbons are running along alternating directions perpendicular to the $c$-axis. 
Here, GeCu$_{2}$O$_{4}$ samples of high quality were prepared via high pressure synthesis (at 4~GPa) and the corresponding magnetic and dielectric properties  were investigated. For the first time, we observed a ferroelectric polarization emerging at T$_{N}$ $\sim$ 33~K.
Although the ferroelectric polarization is weak in GeCu$_{2}$O$_{4}$ ($P$~$\sim$~0.2$\mu$C/m$^{2}$), the existence of spin-induced multiferroicity provides a strong constraint on the possible ground state magnetic structures and/or the corresponding theoretical models of multiferroicity for GeCu$_{2}$O$_{4}$.

\end{abstract}
%\pacs{75.85.+t, 77.70.+a, 77.22.-d,75.25.-j }
\maketitle

%\section{Introduction}

Low dimensional $S = 1/2$ quantum spin systems have been a challenging topic of continued interest within contemporary condensed matter physics. A vast amount of materials with atomic arrangements including quasi two-dimensional (2D) layers based on square lattices, plaquettes, triangular or Kagome lattices, quasi one-dimensional (1D) spin chains or ladders are known to exhibit a wide variety of fascinating quantum phenomena at low temperature \cite{QM} such as superconductivity, spin dimerization, Bose-Einstein condensation, quantum spin liquid states and multiferroicity which can be observed in these systems with competing charge, spin, orbital and lattice degrees of freedom.

Transition metal oxides with spinel structure have the general chemical formula $A$$B$$_{2}$O$_{4}$.
These materials have been a playground for theoretical studies on frustrated quantum magnets. The $B$-ions form a three-dimensional (3D) network of corner-sharing tetrahedra similar to a pyrochlore lattice which gives rise to strong geometrical frustration for antiferromagnetically coupled $B$-site spins. Theoretically, a ground state of quantum spin liquid can be realized in the ideal isotropic pyrochlore lattice. It is well known that many spinel magnetic compounds exhibit unusual magnetic properties and various ground states with or without long range magnetic ordering. The material-specific details like ion anisotropy and structural distortions are able to lift the high degeneracy \cite{FM}.

GeCu$_{2}$O$_{4}$ crystallizes in a strongly distorted spinel structure (Hausmannite) with space group $I4_{1}/amd$  \cite{Hegen}. The lattice parameters amount to $a$~$=$~$b$~$=$~5.593~\AA\ and $c$~$=$~9.396~\AA. GeCu$_{2}$O$_{4}$ exhibits drastically elongated CuO$_{6}$ octahedra in $c$-direction due to a strong Jahn-Teller effect for the Cu$^{2+}$ ions, see Fig.~1(a).
The $ab$-plane "intra-chain" exchange $J$ in $a$- or $b$-direction is much larger than the "inter-chain" exchange $J_{x}$ in $c$-direction.
From the analysis of its magnetic susceptibility $J_{x}/J \sim 0.16$ has been reported    \cite{Yamada}. The interaction between chains in the same layer is also weak. The directions of these chains are orthogonal to these lying in the neighboring layers.

Theoretically the real pyrochlore lattice can be simplified by the projection onto a 2D checkerboard pattern known as the planar pyrochlore antiferromagnetic and crossed-chain model (CCM), which conserving the magnetic frustration and interaction anisotropy. In the simple anisotropic limit ($J_{x} \ll J$) which might be assumed for GeCu$_{2}$O$_{4}$ a crossed-dimmer ground state is predicted  \cite{CCM}.
Later, a more detail work combined with density function calculation proposed the ground state of GeCu$_{2}$O$_{4}$ being a spin spiral state along the chains of edge-sharing CuO$_{4}$ plaquettes  \cite{Tsirlin}. The spiral state is common for many quasi-1D frustrated $J_{1}$-$J_{2}$ spin chain compounds. Such a spiral magnetic structure is able to induce electric polarization and strong magnetoelectric coupling \cite{Seki}. However, the calculation \cite{Tsirlin} predicts an opposite twisting direction for neighboring chains arising from the inter-chain exchange and, thus, no macroscopic polarization.

So far, only a few measurements have been reported on the physical properties of GeCu$_{2}$O$_{4}$   \cite{Yamada, Zou} since the spinel is meta-stable and can only be synthesized at high pressure \cite{Hegen}.
In spite of the three-dimensional nature of the spinel structure, a magnetic susceptibility characteristic for a quasi-1D Heisenberg spin 1/2 chain system with $J$ of $\sim$135~K has been reported  \cite{Yamada}. Moreover, long range antiferromagnetic ordering emerges below 33~K. Very recently, a powder neutron diffraction experiment \cite{Zou} reported a novel collinear 'up-up-down-down' ($\uparrow$$\uparrow$$\downarrow$$\downarrow$) pattern along each spin chain within the $ab$ plane, which contradicts previous theoretic predictions \cite{Tsirlin}. However, also for the experimentally observed magnetic structure \cite{Zou} no multiferroicity can be expected.

\begin{figure}
\includegraphics[width=1\columnwidth]{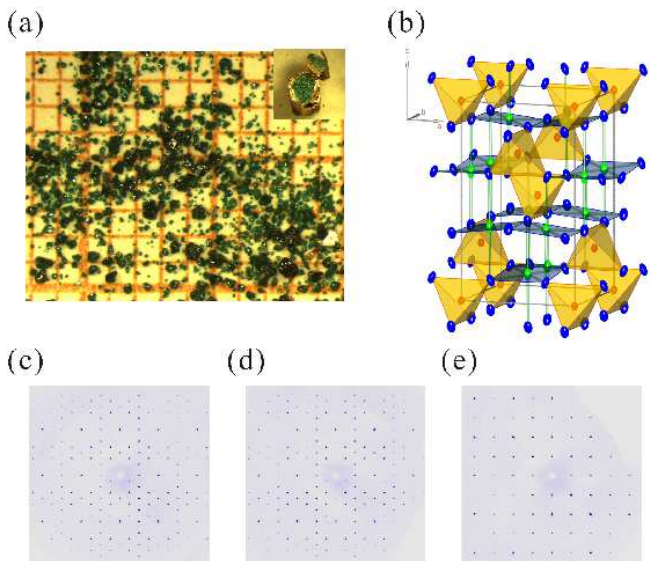}
\caption{
\label{Fig1}
(a) A photo of our synthesized GeCu$_{2}$O$_{4}$ single crystals.
(b) Representation of the crystal structure of GeCu$_{2}$O$_{4}$ as obtained from our single crystal X-ray diffraction measurement. The Cu-oxygen plaquettes are indicated by the blue areas; red/green/blue ellipsoids denote Ge, Cu and O ions (99\%\ probability ellipsoids). 
(c-e) x-ray scattering intensities in the $0$$K$$L$, $H$$0$$L$ and $H$$K$$0$ plane of reciprocal space indicating the single crystalline nature of our GeCu$_{2}$O$_{4}$ single crystal.
}\end{figure}
\begin{table}
\begin{ruledtabular}
\begin{tabular}{l|cccc}
\hline
\textbf{atom} & \textbf{x} & \textbf{y} & \textbf{z}  \\
Ge1 & 0 & 0 & 0   \\
Cu1 & 0 & 1/4 & 5/8   \\
O1  & 0 & 0.24171(9) & 0.35882(6)   \\
\hline
\textbf{atom} & \textbf{U$_{\mathrm{11}}$ (\AA$^2$)} & \textbf{U$_{\mathrm{22}}$ (\AA$^2$)} & \textbf{U$_{\mathrm{33}}$ (\AA$^2$)}    \\
Ge1 & 0.00196(3) & 0.00196(3) & 0.00389(5)     \\
Cu1 & 0.00273(5) & 0.00238(4) & 0.00786(5)    \\
O1  & 0.00367(16) & 0.00306(16) & 0.00725(15)    \\
\hline
\textbf{atom} & \textbf{U$_{\mathrm{12}}$ (\AA$^2$)} & \textbf{U$_{\mathrm{13}}$ (\AA$^2$)} & \textbf{U$_{\mathrm{23}}$ (\AA$^2$)}    \\
Ge1 & 0 & 0 & 0     \\
Cu1 & 0 & 0 & 0.00010(3)    \\
O1  & 0 & 0 & 0.00184(11)     \\
\hline
\textbf{atom} & \textbf{U$_{\mathrm{iso}}$ (\AA$^2$)} & \textbf{occup.}  &     \\
Ge1 & 0.00260(2) & 1  &       \\
Cu1 & 0.00432(3) & 1  &      \\
O1  & 0.00466(9) & 1  &      \\
\hline
\end{tabular}
\end{ruledtabular}
\caption{Structural parameters derived from the crystal structure refinement of our single crystal X-ray diffraction measurement;
$a$~=~5.5929(4)\AA, $c$~=~9.3984(6)\AA, space group: \emph{I4$_1$/amd}.}
\end{table}

\begin{figure}[!h]
\includegraphics[width=0.85\columnwidth]{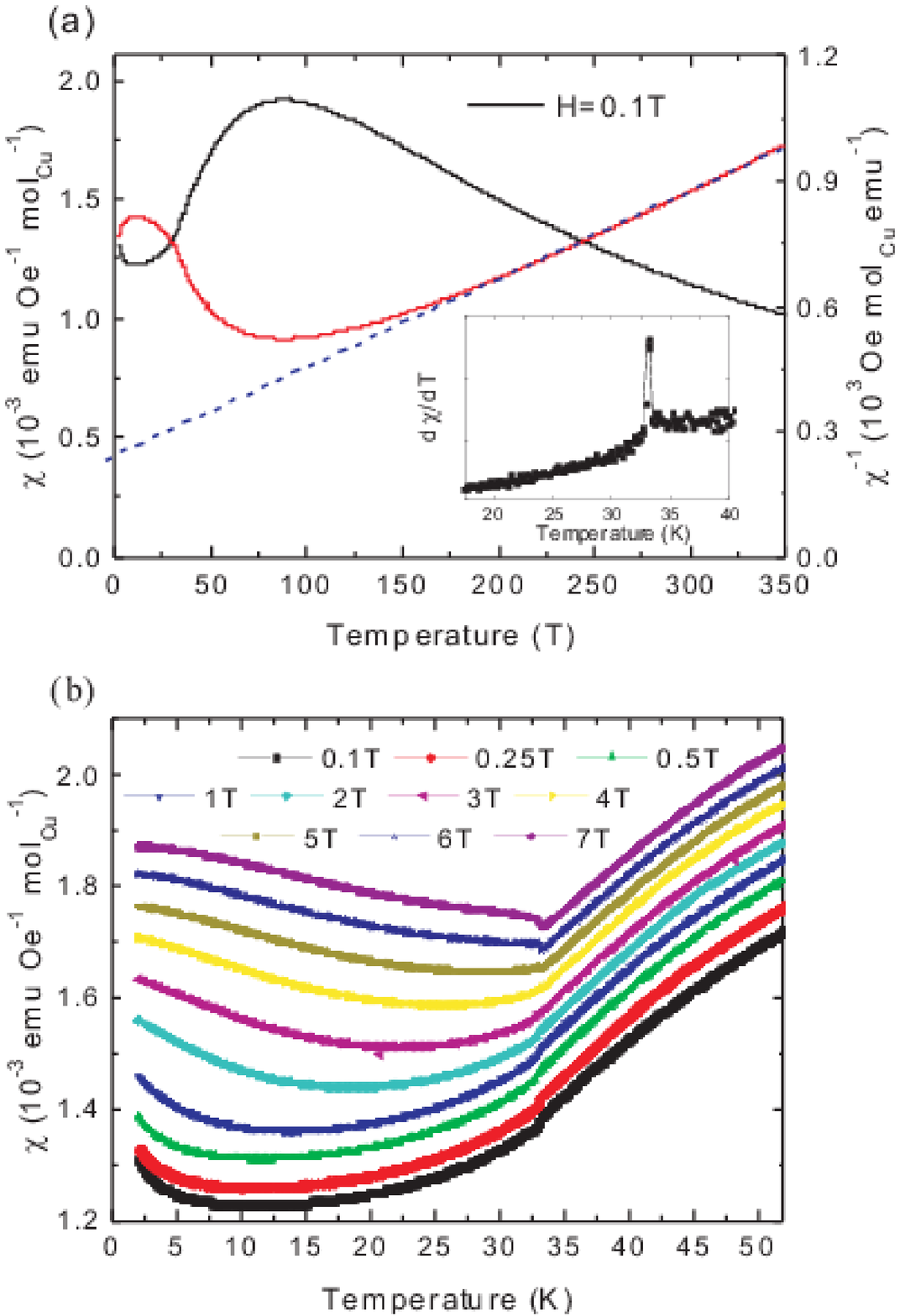}
\caption
{(a) The temperature dependence of the magnetic susceptibility $\chi$(T)  of GeCu$_{2}$O$_{4}$ (black) and $\chi^{-1}$ (red) measured in an external field $H$ of 0.1~T. 
The dashed line denotes a Curie-Weiss fit. In the inset the derivative ($d\chi/dT$) is shown. 
(b) $\chi$ measured under different applied fields $H$ ranging from 0.1~T to 7~T). For clarity, all the curves are shifted vertically except for $H$~=~0.1~T.
}\end{figure}

We noticed that the complex structure of GeCu$_{2}$O$_{4}$ which consists of intertwined quasi-1D CuO$_{2}$ spin chains strongly resembles some structural motifs in CuO$_{2}$Cl$_{2}$ \cite{CCO0}.
CuO$_{2}$Cl$_{2}$ is a newly discovered spin-driven multiferroic material \cite{CCO1} with high transition temperature ($\sim$70~K).
Hence, we synthesized GeCu$_{2}$O$_{4}$ single crystals at high pressures and analyzed the dielectric and pyroelectric properties of this spinel oxide. Our measurements reveal the unexpected existence of multiferroicity in GeCu$_{2}$O$_{4}$ questioning the reported models of its magnetic ground state.

%\section{Experimental}

First, polycrystalline mixtures were prepared by solid-state reaction. The starting materials of GeO$_{2}$ and CuO were mixed in a molar ratio of 1~:~2 and, then, sintered for 50~h at 850$^{\circ}$C in air with intermediate grindings. The resulting blue product is composed of GeCuO$_{3}$ and CuO as confirmed by powder x-ray diffraction. 
Next, the obtained powder was sealed in cylindrical gold capsules, which were inserted in a BN sleeve and, then, pressurized with a multi-anvil high pressure apparatus to 4~GPa pressure and, finally, heated to 900$^{\circ}$C. After annealing at 900 $^{\circ}$C for 30 minutes,
the sample was cooled to room temperature before the high pressure was released. 
Quasihydrostatic pressures were generated with a Walker-type multi-anvil assembly \cite{walker} in combination with a hydraulic press for force generation. For pressure redistribution, we used octahedra manufactured from MgO with 5\%\ Cr$_2$O$_3$ and an edge length of 18~mm. Calibration of p, T conditions was performed independently before the experiments. Temperatures were measured with type-C thermocouples and pressures were determined using the resistivity changes of Bi and Pb at room temperature.
Within the gold capsule a dense green pillar could be obtained after the high pressure synthesis. 
This color change from blue to green has also been observed in previous reports of the high pressure synthesis of GeCu$_{2}$O$_{4}$ \cite{Hegen}.
If the dwelling time at 900 $^{\circ}$C was enhanced and the subsequent cooling rate was decreased, the aggregation of small transparent green 
single crystals (grown via self-nucleation) of 50~$\mu$m to 500~$\mu$m size could be observed, see Fig.~1(a).  

Powder x-ray diffraction (XRD) measurements have been performed using
with Cu K$_{\alpha 1}$ radiation on a \emph{Bruker D8 Discover A25} powder
x-ray diffractometer indicating a single phase of GeCu$_{2}$O$_{4}$. 
%The FullProf program package was used for Le Bail fits and Rietveld refinement. 

Single-crystal XRD measurements have been performed
at room-temperature using Mo K$_{\alpha}$
radiation on a \emph{Bruker D8VENTURE} single-crystal x-ray
diffractometer equipped with a bent graphite monochromator
for about 3$\times$ intensity enhancement and a \emph{Photon} CMOS
large area detector. An as-grown single crystal with roughly 40-$\mu$m
diameter has been measured and a multiscan absorption
correction has been applied to the data (minimum and
maximum transmission: 0.5706 and 0.7536, respectively).
For space group \emph{I4$_1$/amd} 17773 reflections ($H$: −13 $\rightarrow$
10, $K$: −12 $\rightarrow$ 14, and $L$: −21 $\rightarrow$ 21) have been collected
with an internal R value of 2.71\%, a redundancy of 24.0
and 99\%\ (89.7\%) coverage up to $2$$\Theta$ = 117.7$^{\circ}$ (144.5$^{\circ}$).
The Goodness of fit, the R- and weighted R-values within the 
crystal structure refinement amount to 1.63, 1.53\%\ and 4.27\%
respectively.
The structural parameters are listed in Tab.~1 and the obtained crystal structure is plotted in Fig.~1(b).
The x-ray scattering intensities within different planes in reciprocal space are shown in Fig.~1(c-e)
and indicate the single crystalline nature of our measured crystal.

For the study of the dielectric properties of GeCu$_{2}$O$_{4}$, the dense polycrystalline sample was cut and polished to thin plates with thickness of 0.1-0.3~mm. Silver paint was applied to both sides as electrodes to form parallel plate capacitors whose capacitance are proportional to the dielectric constant $\epsilon$. The samples have been mounted on the cryogenic stage and inserted in a \emph{Quantum Design} 9T-PPMS. A high-precision capacitance bridge (\emph{AH 2700A, Andeen-Hagerling Inc.}) was used for dielectric measurements. No apparent difference was observed in the different measuring conditions (various excitation voltages, sweeping rates and integration times have been applied). The electric polarization has been obtained by the pyroelectric current method. 
First, we polarized the specimens during the cooling process with a static electric field of 1-5~MV/m. Then, we short-circuited both sides of the sample at base temperature for about one hour in order to remove any possible trapped interfacial charge carriers. 
The pyroelectric current was measured on heating with a heating rate of 3-5~K/min. 
The electric polarization $P$ has been obtained from the integration of the measured pyroelectric current. Magnetic properties have been measured on a SQUID magnetometer (\emph{MPMS3, Quantum Design Inc.}).

%\section{Experimental Results and Discussions}

\begin{figure}[!h]
\includegraphics[width=0.7\columnwidth]{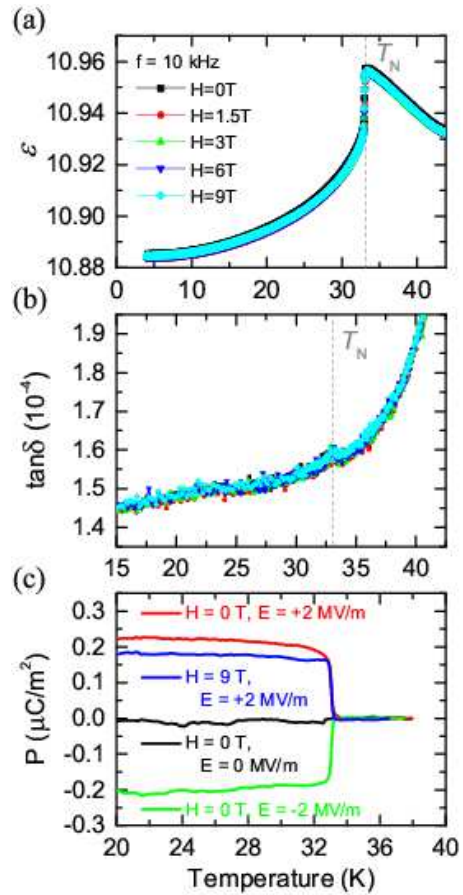}
\caption
{Temperature dependence of the dielectric constant $\epsilon$(T) of polycrystalline GeCu$_{2}$O$_{4}$  under different magnetic fields $H =$ 0~T - 9~T.   (b) The corresponding tangential loss tan$\delta$. All these curves have been measured at the frequency of 10~kHz.
(c) The temperature dependence of the electric polarization (obtained by integration of the pyroelectric current) 
which was measured after different poling processes (+2~MV/m, 0~MV/m and -2~MV/m). % for fields of $H$~$=$~$0$ and 9~T.
}
\end{figure}

The temperature dependence of the magnetic susceptibility $\chi$(T) of GeCu$_{2}$O$_{4}$ measured in a field of 0.1~T is shown in Fig.~2(a). Zero-field-cooled (ZFC) and field-cooled (FC) measurements  coincide. Around 80~K a broad hump appears  which is characteristic for short range antiferromagnetic ordering due to frustration effects. A simple Curie-Weiss fit within the temperature range of 180~K to 350~K) yields a Curie-Weiss temperatures $\theta_{CW}$ of about -110~K and an effective magnetic moment $\mu_ {eff}$ which amounts to $\sim$1.94~$\mu$B which is similar with previous studies \cite{Yamada}. 
The value of $\mu_{eff}$  is slightly higher than that might be expected for the paramagnetic Cu$^{2+}$ spin-only value (1.73~$\mu B$).
At low temperature $\chi$(T) exhibits a slight drop at $T_{N}$~$\sim$~33~K suggesting the emergence of long range magnetic ordering. As shown in the inset,  the corresponding temperature derivative exhibits a sharp peak at $T_{N}$.

We also studied the magnetic field dependence of $\chi$(T) for different fields up to 7~T. 
All $\chi$(T) curves coincide above $T_{N}$ (including the value of $T_{N}$). On cooling below $T_{N}$, the temperature dependence of $\chi$(T) changes from a decreasing behavior in low fields to an increasing one in high fields above 4~T.

Our results resemble the ones reported in Ref.~\cite{Zou} where $\chi$(T) was also explained by adding a minor contribution 
of paramagnetic impurities like GeO$_{2}$. In fact, about 3.6\%\ GeO$_{2}$ was found in their samples according to x-ray diffraction.  However no such impurities exist in our GeCu$_{2}$O$_{4}$ samples. 
(We also measured the magnetic susceptibilities of pure GeO$_{2}$, GeCuO$_{3}$, and CuO powder as reference materials in order to search for possible impurities below the detection level of x-ray diffraction. But, no anomalies can be detected around 33~K for these reference materials, confirming that our observed field evolution of $\chi$(T) is intrinsic for GeCu$_{2}$O$_{4}$.)

The temperature-dependence of the dielectric constant $\epsilon$(T)  measured at a frequency of 10~kHz for different fields $H$ is shown in Fig.~3(a). 
On cooling $\epsilon$ slightly increases for temperatures somewhat above $T_{N}$  and, then, $\epsilon$ drops sharply at the magnetic transition temperature T$_N$. In our experiments, we also measured $\epsilon$ with various frequencies between 1~kHz and 20~kHz (not shown here), and, the observed dielectric anomaly shows no difference for all tested measuring frequencies. This confirms the intrinsic origin of the dielectric anomaly at T$_N$ which -although weak- can also be seen in the corresponding dielectric loss, see Fig.~3(b). As shown in Fig.~3(a-b), the effect of external magnetic fields on the dielectric behavior is almost negligible for the whole studied range of magnetic fields $H$ up to 9~T.  

Moreover, we measured also the pyroelectric current in order to study the electric polarization of GeCu$_{2}$O$_{4}$. 
A small but sharp pyroelectric current peak emerges at $T_{N}$. The electric polarization in GeCu$_{2}$O$_{4}$ was obtained by integrating the small pyroelectric currents with our high SNR (Signal Noise Ratio) setup, see Fig.~3(c). 
To minimize the accumulative error and uncertainty due to noise within the background, the integration was carried out from above $T_{N}$  to just above 20~K where the polarization has almost saturated ($P_{s}$ in our notion). 
Although $P_{s}$ is quite small -about 0.2 $\mu$C/m$^{2}$ at 20~K- it can be inverted with opposite poling
whereas the measured pyroelectric current and corresponding polarization is negligible for zero poling field, see Fig.~3(c).
Our observation reveals the weak ferroelectric nature of GeCu$_{2}$O$_{4}$ below $T_{N}$. GeCu$_{2}$O$_{4}$, the polarization and magnetoelectric effect is much weaker than in most known multiferroic cuprates as LiCu$_{2}$O$_{2}$  \cite{LiCu2O2}. 
However, for the first time, we discovered a new multiferroic material with a strongly tetragonally distorted spinel structure.  

The appearance of an ferroelectric polarization at the magnetic ordering temperature is indicative for spin-induced multiferroicity in GeCu$_{2}$O$_{4}$.
The appearance of an electric polarization that is induced by the magnetic structure has been studied thoroughly for the last decade \cite{Cheong, Tokura, Wang}. Three major microscopic origins of spin-induced ferroelectricity are known:  (i) exchange-striction arising from the symmetric spin exchange interaction in certain collinear spin ordering states in some frustrated systems with magnetically inequivalent ions. The "`up-up-down-down"' spin arrangement along the atomically alternating A-B spin chains observed in Ca$_{3}$(Co,Mn)O$_{6}$ \cite{CCMO} is an example for this first mechanism;
(ii) the Inverse Dzyaloshinskii-Moriya (DM) or spin current model. Via spin-orbit interaction two non-collinear magnetic moments are able to induce a local electric polarization ${\bf P}_{ij}=A \hat{\bf {e}}_{ij} \times ( {\bm S}_i  \times {\bm S}_j )$  with   ($\hat{\bf {e}}_{ij} $ being the unit vector between the two sites $i$ and $j$.); and (iii) the Arima model where the spin-dependent $p$-$d$ hybridization induces a local polarization   $P_{il}\propto(S_{i}\cdot\hat{e}_{il} )^{2}\hat{e}_{il} $  with  ($\hat{e}_{il} $ being the unit vector connecting magnetic ion and ligand site.  
The last two models (ii) and (iii) require certain magnetic and/or lattice structure such that the sum over the crystal lattice sites is not canceled out and such that a non-zero macroscopic polarization $P$ emerges. Model (ii) is able to yield non-zero $P$ e.g. for cycloidal magnetic structures \cite{LiCu2O2} whereas model (III) requires a certain Bravais lattice that hosts e.g. a screw-type helical magnetic structure \cite{Arima}. For both models (ii) and (iii) the magnitude of the induced polarization $P$ usually rather weak.

In GeCu$_{2}$O$_{4}$ the presence of identical magnetic ions -i.e. Cu$^{2+}$ ions- excludes the possibility of mechanism (i). 
The lattice symmetry (space group I4$_1$/amd) also excludes the possibility of the last mechanism (iii). 
For the spiral magnetic ground state that was theoretically predicted  \cite{Tsirlin} the spins are spiraling along each chain with the renormalized pitch angle of about 84$^{\circ}$. But, the induced polarization cancels out due to opposite twisting directions in neighboring chains.  
Also for the cross-dimer state \cite{CCM} no polarization $P$ can be expected since this dimer ground states does not break inversion symmetry. 
For example, GeCuO$_{3}$ is such a spin-dimer system where no dielectric anomaly could be observed at $T_{N}$ \cite{GeCuO3}.
So far, there is only one reported neutron diffraction measurement on a GeCu$_{2}$O$_{4}$ powder sample \cite{Zou}. 
The experimentally measured collinear "`up-up-down-down"' spin structure is also centrosymmetric and incompatible with emergence of spin-induced multiferroicity. Hence, further studies on its spin structure based on high quality powder samples or single crystals are necessary for an understanding of the mechanism of multiferroicity in GeCu$_{2}$O$_{4}$.  

%\section{Summary}

Conclusively, in this work, we presented magnetic, dielectric and polarization measurements on the metastable GeCu$_{2}$O$_{4}$ spinel oxide prepared by high pressure synthesis. We observed spin-induced ferroelectricity in GeCu$_{2}$O$_{4}$, where all the experimentally or theoretically proposed spin structures are not compatible with existing theoretical models for multiferroicity (or vice versa). 
Due to the complex intertwining of quasi-1D CuO$_{2}$ spin chains and strong inherent frustration effects within the tetragonally distorted spinel structure, 
interesting physics can be expected as an outcome of future theoretical and experimental studies on this new multiferroic material.

\begin{acknowledgments}
We would like to thank R. Castillo for the support in the sample preparation.
We also would like to thank L.~H.~Tjeng for his support and for helpful discussions.
The research is partially supported by the Deutsche Forschungsgemeinschaft through SFB 1143 and through DFG project number 320571839.
\end{acknowledgments}

\end{document}